\input harvmac
\input psfig
\newcount\figno
\figno=0
\def\fig#1#2#3{
\par\begingroup\parindent=0pt\leftskip=1cm\rightskip=1cm\parindent=0pt
\global\advance\figno by 1
\midinsert
\epsfxsize=#3
\centerline{\epsfbox{#2}}
\vskip 12pt
{\bf Fig. \the\figno:} #1\par
\endinsert\endgroup\par
}
\def\figlabel#1{\xdef#1{\the\figno}}
\def\encadremath#1{\vbox{\hrule\hbox{\vrule\kern8pt\vbox{\kern8pt
\hbox{$\displaystyle #1$}\kern8pt}
\kern8pt\vrule}\hrule}}
\def\underarrow#1{\vbox{\ialign{##\crcr$\hfil\displaystyle
 {#1}\hfil$\crcr\noalign{\kern1pt\nointerlineskip}$\longrightarrow$\crcr}}}
%
\overfullrule=0pt

%
\def\C{{\bf C}}
\def\tilde{\widetilde}

\def\S{{\bf S}}

\font\zfont = cmss10 

\def\bigone{\hbox{1\kern -.23em {\rm l}}}
\def\ZZ{\hbox{\zfont Z\kern-.4emZ}}

\Title{hep-th/0106109}
{\vbox{\centerline{Quantum Gravity In De Sitter Space}}}
\smallskip
\centerline{Edward Witten}
\smallskip
\centerline{\it Dept. of Physics, Cal Tech, Pasadena, CA}
\smallskip
\centerline{and}
\smallskip
\centerline{\it CIT-USC Center For
Theoretical Physics, USC, Los Angeles CA}

\medskip

\noindent

We discuss some general properties of quantum gravity in de Sitter
space.  It has been argued that the Hilbert space is of finite
dimension.  This suggests a macroscopic argument that General Relativity
cannot be quantized -- unless it is embedded in a more complete
theory that determines the value
of the cosmological constant.
We give a definition of the quantum Hilbert space 
using the asymptotic behavior in the past and future, without
requiring detailed microscopic knowledge.
We discuss the difficulties in defining any precisely calculable or
measurable observables in an asymptotically de Sitter spacetime, and
explore some meta-observables that appear to make mathematical sense
but cannot be measured by an observer who lives in the spacetime.
This article is an expanded version of a lecture at Strings 2001 in Mumbai.

\Date{June, 2001}

De Sitter $n$-space or ${\rm dS}_n$ is the maximally symmetric 
$n$-dimensional spacetime with positive
cosmological constant $\Lambda$.  
Its symmetry group is $SO(1,n)$.  If we introduce variables $x_0,x_1,
\dots, x_n$ obeying $x_0^2-\sum_{1=1}^nx_i^2$, the de Sitter metric
is simply (up to a constant factor)
\eqn\dsone{ds^2=-dx_0^2+\sum_{i=1}^ndx_i^2.}
Alternatively, one can write the metric as
\eqn\dsmetric{ds^2=-dt^2+\cosh^2t \,d\Omega^2,}
where $d\Omega^2$ is the metric on a unit round $(n-1)$-sphere. 
This spacetime has compact spatial sections (such as $t=0$), so when
we speak of {\it asymptotically} de Sitter space -- as we should in the
presence of gravity, since the metric fluctuates -- the asymptopia in
question is in the past and future.  There is no notion of spatial infinity.
This is in sharp contrast with
Anti de Sitter space, the maximally symmetric spacetime of negative
cosmological constant, where, as we have come to know well in the last
few years, asymptopia is at spatial infinity.  It also contrasts with
Minkowski space, which from a conformal point of view has a natural
null infinity.

In de Sitter space, there is no positive conserved energy.  In fact,
no matter what generator we pick for $SO(1,n)$, the corresponding Killing
vector field, though perhaps timelike in some region of de Sitter space,
is spacelike in some other region.  
For example, a typical Lorentz generator in de Sitter space is
\eqn\typgen{K=x_1{\partial\over \partial x_0}+x_0{\partial\over \partial x_1}.}
Whether this generator moves us forwards or backwards in time
(towards increasing or decreasing $x_0$) depends on the sign of $x_1$.  The conserved charge
associated with $K$ is positive for excitations supported at positive
$x_1$ and negative for those at negative $x_1$.  This is the best we can
do: there is no positive conserved energy in de Sitter space.

Consequently, there cannot be unbroken supersymmetry in de Sitter space.
If there is a nonzero supercharge $Q$, we can (possibly after replacing
$Q$ by $Q+Q^\dagger$ or $i(Q-Q^\dagger)$) assume that $Q$ is Hermitian.
Then $Q^2$ cannot be zero, and is a nonnegative bosonic conserved quantity;
but there is no such object.

We can rotate de Sitter space to Euclidean signature by setting
$x_0\to ix_0$ (or equivalently, set
$t=i\tau$ and take $\tau=\pi/2-\theta$).  The Euclidean continuation
is a standard $n$-sphere $\S^n$, with symmetry group $SO(n+1)$.
After the continuation, the operator $K$ becomes the generator of
a rotation, and obeys
\eqn\inovo{\exp(2\pi K)=1.}
Because of this, the Euclidean de Sitter path integral can be interepreted
in terms of a thermal ensemble.
This leads to the notion of a de Sitter temperature
\ref\fhn{R. Figari, R. Hoegh-Krohn, and C. R. Nappi,
``Interacting Relativistic Boson Fields in The De Sitter Universe
With Two Spacetime Dimensions,'' Commun. Math. Phys. {\bf 44}
(1975) 265.} and the associated entropy
\ref\gibhawk{G. W. Gibbons and S. W. Hawking, ``Cosmological Event
Horizons, Thermodynamics, and Particle Creation,''
Phys. Rev. {\bf D15} 2738 (1977).}.  
Like the Bekenstein-Hawking entropy of a black hole, the de Sitter
entropy can be written
\eqn\polo{S={A\over 4G},}
where $G$ is Newton's constant, and $A$ is the area of a horizon.
In this case, however, the horizon is observer-dependent, and because
of this it is not entirely clear which concepts about black holes
carry over to de Sitter space.

An observer in de Sitter space can only see part of the space.  This is
because of the exponential inflation that occurs in the future: the
space expands so fast that light rays do not manage to propagate
all the way around it.  To make the causal structure
of de Sitter space clear, one can introduce a new ``time'' coordinate $u$ by 
\eqn\normform{u=2\tan^{-1}e^t,}
 so that for $-\infty< t <\infty$, $u$
ranges over $0<u<\pi$.  The metric becomes
\eqn\finform{ds^2={1\over \sin^2u}\left(-du^2+d\Omega^2\right).}
The asymptotic past ${\cal I}_-$ consists of a copy of $\S^{n-1}$ at $u=0$,
and the asymptotic future ${\cal I}_+$ consists of a copy of $\S^{n-1}$
at $u=\pi$.  Any trajectory in de Sitter space begins at some point $P$
in ${\cal I}_-$ and ends at some point $Q$ in ${\cal I}_+$.  From a causal
point of view, in a sense considered by Bousso \ref\bousso{R. Bousso,
``A Covariant Entropy Conjecture,'' JHEP {\bf 9907} (1999) 004, 
hep-th/9905177, ``Holography In General Spacetimes,'' JHEP {\bf 9906}
(1999) 028, hep-th/9906022, ``Positive Vacuum Energy And The $N$-Bound,''
hep-th/0010252.},
any observer can be identified with the pair $(P,Q)$.
The region of de Sitter space that one can influence, and likewise the
region that one can see, depend only on $P$ and $Q$, and not
on the details of one's trajectory in spacetime.  
What one can see is determined only by $Q$, and the region that one
can influence depends only on $P$.  

To describe in detail the horizon of an observer, let us write
$d\Omega^2=d\chi^2+\sin^2\chi d\tilde \Omega^2$, where $\chi$ is a polar
angle,
ranging over $0\leq \chi\leq \pi$, and $d\tilde\Omega^2$
is the round metric on an $(n-2)$-sphere.  The de Sitter metric
then becomes
\eqn\ormform{ds^2={1\over \sin^2 u}\left(-du^2+d\chi^2+\sin^2\chi \,d\tilde 
\Omega^2\right).}
Consider now an observer who sits at the ``north pole'' of the
sphere, that is, at $\chi=0$.  (In fact, any geodesic in de Sitter
space is equivalent to $\chi=0$ by the action of the de Sitter group.)
From the form of the metric, we see that the propagation of light rays
is bounded by $|d\chi/du|\leq 1$.  Since the spacetime ``ends'' in this coordinate system at $u=\pi$,
a light ray emitted at $\chi>\pi - u$ will never reach the observer at $\chi=0$.
So the boundary of the region that this observer can see is given by
\eqn\tugu{\chi = \pi - u.}
This is the horizon.  In general, the $(n-2)$-sphere of given $\chi$ and
$u$ has metric $(\sin\chi/\sin u)^2d\tilde\Omega^2$, and its area
is proportional to $(\sin\chi/\sin u)^{n-2}$.  Relating $\chi$ to
$u$ by \tugu, we see that the horizon area is time-independent.  This
is in keeping with general theorems saying that the area of the past
horizon of an observer cannot decrease in time.  For de Sitter space,
this horizon area is precisely constant, and for a generic perturbation
of de Sitter space, it is an increasing function of time.

By studies of $D$-branes and in a variety of other ways,
we have learned in the last few years
to interpret the Bekenstein-Hawking entropy of a black
hole like every other entropy in statistical mechanics: it is the logarithm
of the number  of quantum states of the black hole.
It has been argued \ref\banks{T. Banks, ``Cosmological Breaking Of
Supersymmetry, or Little Lambda Goes Back To The Future II,''
hep-th/0007146.} that the same holds for de Sitter space, more
precisely 
that the Hilbert space of quantum gravity in
asymptotically de Sitter space time has a finite dimension $N$,
and that the entropy of de Sitter space is 
\eqn\enform{S=\ln \,N.} 
This is a very interesting answer for many reasons, including the fact
that the de Sitter Hilbert space has infinite dimension perturbatively.
There is no contradiction here, since perturbation theory 
is an expansion in powers of $G\Lambda^{(n-2)/n}$. If the above
formula for $N$ is correct, then $N$ diverges exponentially as $G\Lambda^{(n-2)/n}\to 0$, so perturbation theory, in exhibiting
an infinite-dimensional Hilbert space, gives the right answer
for the weak coupling limit.
Moreover, one can argue to a reasonable extent that perturbation theory
may be breaking down in this situation as one moves away from the weak
coupling limit.  (The arguments have been explained
to me by T. Banks, R. Bousso, and G. Horowitz, and I will briefly allude
to them later.)
However, there is no known controlled calculation incorporating the breakdown
of perturbation theory and exhibiting the claimed behavior of $N$.

If the quantum Hilbert space in de Sitter space really has a finite
dimension $N$, this gives a strong hint that Einstein's theory with Lagrangian
\eqn\lollipop{I=-{1\over 8\pi G}\int d^nx\sqrt g R -\Lambda \int d^nx \sqrt g}
cannot be quantized and must be derived from a more fundamental theory that determines the possible values of $G\Lambda^{(n-2)/n}$.  Indeed, $N$ is
claimed to be a nontrivial function of $G\Lambda^{(n-2)/n}$ (since it
diverges exponentially as $G\Lambda^{(n-2)/n}\to 0$) yet it obviously takes
integer values.  Clearly, this implies that
 $N$ cannot be a continuous function of
$G\Lambda^{(n-2)/n}$ if the latter can vary continuously.  This suggests that Einstein's theory cannot be
quantized for general values of $G$ and $\Lambda$
but must be derived from a more fundamental theory that
determines the possible values of $G\Lambda^{(n-2)/n}$.

This conclusion should not seem too surprising, since a similar argument
-- at least for some values of $n$ -- can be made for negative cosmological
constant.  
In that case, quantum gravity in asymptotically Anti de Sitter space
is related to a conformal field theory on the boundary.  For $n=3$,
the boundary theory has central charge $c$ proportional to $G\Lambda^{1/3}$
(as was seen from a canonical point of view in \ref\bfh{J. D. Brown
and M. Henneaux, ``Central Charges In The Canonical Realization Of Asymptotic
Symmetries: An Example From Three-Dimensional Gravity,'' Commun. Math.
Phys. {\bf 104} 207 (1986).} and later understood as a special case of the
AdS/CFT correspondence).  The Zamolodchikov $c$-theorem implies that
$c$ is constant in a family of  conformal field theories in
two dimensions,\foot{To be more precise, $c$ is constant in a family of
unitary theories with normalizable vacuum.  These properties are
expected to hold for CFT's arising in the AdS/CFT correspondence.} so $G\Lambda^{1/3}$ cannot be continuously
varied. The same argument holds in any dimension $n>3$ for which there is a suitable
$c$-theorem.

Going back to de Sitter space, what values of $N$ are in fact possible?
If classically there existed compactifications to de Sitter space
(perhaps depending on discrete fluxes to introduce an integer), then in the
classical limit we would have $N\to\infty$.   However, an important no
go theorem 
\nref\dewsh{B. de Wit, D. J. Smit, and N. D. Hari Dass, ``Residual
Supersymmetry Of Compactified $D=10$ Supergravity,'' Nucl. Phys. {\bf B263}
(1987) 165.}%
\nref\maldanun{J. Maldacena and C. Nunez, ``Supergravity
Description Of Field Theories On Curved Manifolds And A No Go Theorem,''
hep-th/0007018, Int. J. Mod. Phys. {\bf A16} 822 (2001).}%
\refs{\dewsh,\maldanun} says 
that there is no classical compactification of ten- or eleven-dimensional
supergravity
to de Sitter space of any dimension.  This means that there is
no classical way to get de Sitter space from string theory or $M$-theory. By a ``classical'' compactification,
I would mean a family of compactifications in which $G\Lambda^{(n-2)/n}$
becomes arbitrarily small and a supergravity or string theory description
becomes arbitrarily good.  The no go theorem means that this does not exist.

In fact, classical or not, I don't know  any clear-cut way to get de Sitter space
from string theory or $M$-theory.  This last statement is not very surprising
given the classical no go theorem.
For, in view of the usual problems in stabilizing moduli, it is hard
to get de Sitter space in a reliable fashion at the quantum level given
that it does not arise classically. (For an analysis of a situation
in which most moduli can be stabilized, leading in the large volume
limit to a nonsupersymmetric vacuum with $\Lambda=0$, see
\ref\poletal{S. B. Giddings, S. Kachru, and J. Polchinski,
``Hierarchies From Fluxes In String Compactifications,''
hep-th/0105097.}.)

The absence of a classical de Sitter limit suggests that the possible
values of $N$ in string/$M$-theory are sporadic, rather than arising
from infinite families, and that there might be only finitely many choices.
If the number of choices is finite, I would not personally expect it to
be possible to get $N>10^{10^{100}}$.  But de Sitter space with such
large $N$ is needed to agree with the most obvious interpretation of recent
astronomical data!\foot{I should note that the point of view of Banks
\banks\ is rather different.   He argues
 that string/$M$-theory as we know it is the Hilbert space
version of a theory that can also be formulated with matrices of
arbitrary finite size $N$.  He takes $N$ as an input and proposes that,
taking into account some conjectured ``large graviton'' effects that cannot
be seen in classical field theory,
some form of $M$-theory dynamics would be possible for any arbitrarily big $N$
and would lead to a scale of supersymmetry breaking that is very low
compared to the Planck mass, but very large compared to the scale
set by the cosmological constant. }

The fact that $N$, the dimension of the Hilbert space ${\cal H}$, is finite means
that the de Sitter symmetry group $SO(1,n-1)$ cannot act
on ${\cal H}$.  Indeed, $SO(1,n-1)$ has no (non-trivial) finite-dimensional
unitary representations!  

This may sound like a problem, but in fact it is not.  The de Sitter
group does not act on ${\cal H}$ because the spatial sections of de Sitter
space are compact.  Always, in General Relativity, the spacetime symmetry
generators (being gauge charges) can be expressed as surface terms at
infinity.  In the case of de Sitter space there is no (spatial) infinity
and hence the de Sitter generators are zero.  Thus, what I have
informally called ``quantum gravity in de Sitter space'' does not
have the invariance of classical de Sitter space.  This is one consequence
of the fact that the only asymptopia is in the past and future.

Absence of de Sitter symmetries should lead one to ask in what precise
sense one might refer to some hypothetical system as ``quantum gravity in
de Sitter space.''  Let us think about the analogous question for 
$\Lambda$ vanishing or negative. It has been clear
since very early studies of quantum gravity \ref\dewitt{B. S. DeWitt,
Quantum Theory Of Gravity II: The Manifestly Covariant Theory,'' Phys.
Rev. {\bf 162} (1967) 1195.} that one cannot define local field operators
in a theory with gravity, but that in an asymptotically flat spacetime of
dimension $n>3$ one can define an $S$-matrix that apparently has all the usual properties.\foot{For $n\leq 3$, the presence of any mass causes the
universe not to be asymptotically flat, and thwarts the existence
of an $S$-matrix in any standard sense.
Above four dimensions, the $S$-matrix acts between Fock states with finitely
many initial and final particles; in four dimensions, because of the usual
infrared problems with massless gauge particles, one must consider
more subtle initial and final states.}  It is reasonable to insist
that a quantum gravity theory in $n$ dimensions for $n>3$ is characterized
by having a unitary, analytic $S$-matrix with a massless spin two
particle that has the expected sort of low energy interations.
This is a satisfactory description of the ``output'' that
a quantum gravity theory in asymptotically flat spacetime should produce,
but  leaves completely open the big question of
how the $S$-matrix is supposed to be computed at least in principle, that is,
what kind of theory produces this output. (Matrix theory
\ref\bfss{T. Banks, W. Fischler, S. H. Shenker, and L. Susskind,
``$M$ Theory As A Matrix Model: A Conjecture,'' he--th/9610043, Phys. Rev.
{\bf D55} (1997) 5112.} supplies a possible answer in certain situations, though
it has not yet been put in a systematic framework.)  

For $\Lambda<0$, the situation is much better.  There is no notion of
an $S$-matrix in asymptotically Anti de Sitter spacetime, but instead one
has the correlation functions of the boundary conformal field theory.
The local operator product relations of CFT makes the boundary CFT for
$\Lambda<0$ a much richer structure than the $S$-matrix for $\Lambda=0$.
 As $\Lambda$ approaches
zero from below, the boundary CFT degenerates to the $S$-matrix, whose
structure is much less rich.  The degeneration occurs because for a connected correlation
function $\langle \phi_1(x_1)\phi_2(x_2)\dots \phi_n(x_n)\rangle$ of the
boundary CFT to have a limit as $\Lambda$ approaches zero from below,
one must take some of the $x_i$ to the past and some to the future;
what survive are the $S$-matrix elements of the $\Lambda=0$ theory.
In this limit, the local operator product relations of the CFT are lost;
operators creating in states (or out states) just commute or anticommute
with each other.  To my thinking, boundary CFT  for $\Lambda<0$
perhaps can be regarded as not just an output but
a dynamical principle that defines what
we mean by quantum gravity for $\Lambda<0$.  
From this point of view, what is missing is an understanding of what
sort of limit corresponds to having a macroscopic spacetime in the interior.

What are we to make of the ``holography'' of 't Hooft and Susskind?
For $\Lambda<0$, this can be the assertion that quantum gravity is
holographically dual to a conformal field theory on the boundary.
For $\Lambda=0$, the situation is much less satisfactory.
Hopefully, for $\Lambda=0$,
holography means more than a mere assertion that quantum gravity
in asymptotically flat spacetime does not have local field operators
but only an $S$-matrix.  This degree of understanding predates even
the early beginnings of string theory \dewitt.  Hopefully, holography
means that, in some yet unclear way, the theory can be described
(and the $S$-matrix computed)
in terms of degrees of freedom that ``live'' at infinity.  Regrettably,
however, for $\Lambda=0$ it is not very clear how even to begin
in that direction, as Minkowski space
has a natural null infinity (leading back to the notion of an $S$-matrix,
perhaps), but not a natural spatial infinity.\foot{Or it has a spatial
infinity consisting of only one point, which is not very helpful.
For a little more
detail on some issues discussed in the last few paragraphs, see my
talk at Strings '98: http://online.itp.ucsb.edu/online/strings98/witten/.}

At any rate, getting back to the case of $\Lambda>0$, we would ideally
like to understand a dynamical principle generating quantum gravity
theories in asymptotically de Sitter spacetime, but at a very minimum
we would like a reasonable description of what sort of output characterizes
them.  It is not enough to merely have a finite dimensional Hilbert space!
I will not really have an answer to propose, but it does seem that any answer,
if there is one, would have to make use of the behavior in the
asymptotic regions in the far past and future since that is the only
asymptopia we have.

\bigskip\noindent{\it The De Sitter Hilbert Space}

Now we will make a few remarks about the Hilbert space of de Sitter
space.  We start with some standard observations about perturbation theory,{}
\foot{For more on perturbation theory in de Sitter space from
 different points of view, see 
\ref\bda{N. D. Birrell and P. C. W. Davies, 
{\it Quantum Fields In Curved Space} (Cambridge University Press, 1982).}
and
\ref\woodard{N. C. Tsamis and R. P. Woodard,
``The Quantum Gravitational Back-Reaction On Inflation,''
hep-ph/9602316.}.}     and
then we will try to suggest a nonperturbative definition of a Hilbert space.

In perturbation theory, the starting point is a free field in
de Sitter space.  Such a free field can be quantized to obtain
a quantum Hilbert space.  Though there is no notion of a state of
minimum energy, the Hilbert space of a free field does contain
a distinguished
de Sitter-invariant state, which one might call the vacuum, $|\Psi\rangle$.
For a free field,
$|\Psi\rangle$ is the unique quantum state which is de Sitter invariant
and {\it Gaussian}.

Concretely, let $\phi$ be a scalar field, and let $T$ be the spatial
section of de Sitter space obtained by setting $x_0=0$ in \dsone.
Thus, $T$ is the ``equator'' in the Euclidean form $\S^n$ of de Sitter space.
We denote a configuration of the $\phi$-field on $T$ as $\phi(x)$.
A quantum wavefunction, in the ``coordinate space'' representation,
is a functional $\Psi(\phi(x))$.  A Gaussian wavefunction has the form
\eqn\onon{\Psi(\phi)=\exp\left(-{1\over 2}\int_{T\times T}d^{n-1}x\,
d^{n-1}y\,D(x,y)\phi(x)\phi(y)\right)}
with some kernel $D(x,y)$.  (Here $d^nx$ is short for the standard
Riemannian measure $dx_1\dots dx_n\sqrt g$.)
There is a unique  $D(x,y)$ for which a wavefunction of this
form is de Sitter invariant.  

In fact, a Gaussian state can be usefully
characterized by the two-point function $G(x,y)=\langle\Psi|\phi(x)
\phi(y)|\Psi\rangle$.  As usual for Gaussian integrals, $G$ can be obtained
by inverting the operator corresponding to $D$.  
The appropriate de Sitter-invariant $G$ can be readily found by working
in Euclidean signature and regarding $G$ as a function on
$\S^n\times \S^n$ which is then restricted to $T\times T$.  In fact,
$G$ is determined by the standard equation for the propagator
\eqn\ononb{\left(-\Delta^2+m^2+\alpha R\right)G(X,Y)=\delta^n(X,Y).}
Here $X,Y$ denote points in $\S^n$, and we have included a coupling
to the Ricci scalar $R$ as well as a mass term.  Eqn. \ononb\ uniquely
determines $G$ and hence (upon taking an operator inverse) it uniquely
determines a de Sitter-invariant Gaussian state $|\Psi\rangle$.  

In fact,
$|\Psi\rangle$ can conveniently be computed by a path integral 
on a hemisphere $H$  of boundary $T$, say the hemisphere $x_0<0$.
One simply carries out a path integral over fields on  $H$
whose boundary values are given by $\phi$. We 
let ${\cal A}_\phi$ be the space of fields $\Phi$ on $H$ whose
 restriction to $T$ is equal to $\phi$.  Then we define
\eqn\ononx{\Psi(\phi)=\int_{\cal A_\phi} D\Phi\,e^{-I(\Phi)}.}   
With this definition, the two point function $\langle\Psi|\phi(x)\phi(y)
|\Psi\rangle$ can be computed by a path integral on the full sphere
$\S^n$  -- the path integral on one hemisphere computes $|\Psi\rangle$
and the path integral on the other hemisphere computes $\langle \Psi |$.
The correlation  $\langle\Psi|\phi(x)\phi(y)
|\Psi\rangle$ is thus computed, from this point of view, by a path
integral on the full sphere (with $\phi(x)$ and $\phi(y)$ inserted
to the equator), and this makes it clear that it obeys the
usual covariant equation \ononb.

In perturbation theory, other quantum states are polynomials times the
``vacuum'' state $\Psi$.  Thus, they take the general form
\eqn\lonon{\Psi_f(\phi)=
\int dx^{(1)}\dots dx^{(s)}\,\,
f( x^{(1)},\dots,  x^{(s)}) \phi(x^{(1)})\dots \phi(x^{(s)})\,\Psi(\phi).}
(Here $ x^{(1)},\dots,  x^{(s)}$ are $s$ points in $T$.)
If we are doing quantum gravity, we must discard most of these states:
physically acceptable states are invariant under the de Sitter group.
De Sitter invariance gives a severe restriction on $f$.
But as $f$ depends on an arbitrarily large number of points in
$T$, and the de Sitter group has finite dimension,
de Sitter invariant $f$'s do exist for every sufficiently large $s$. Moreover,
for sufficiently large $s$, the space of such $f$'s is infinite-dimensional.
Thus, the perturbative
Hilbert space has infinite dimension. 

So far we have just given a free field description, which could be
used in principle (modulo the usual problems with renormalizability)
as the starting point in constructing perturbation theory.
It is clear that if $s$ is very large, or for fixed $s$ if $f$ is
very rapidly varying, perturbation theory may break down because of
a large gravitational back-reaction.  This is why the fact that the
Hilbert space is infinite-dimensional in perturbation theory does not
guarantee that it is really infinite-dimensional. 
But it is not at all clear how
to actually do a calculation exhibiting the alleged finite-dimensionality
of the quantum Hilbert space ${\cal H}$.  

\bigskip\noindent{\it Nonperturbative Definition of ${\cal H}$}

In fact, our goal here will be far more modest.  We will just try to
give a natural definition of  ${\cal H}$ 
and thus give a precise framework in which, in principle, it might
be possible to eventually address the issue of its finite-dimensionality.

To give a nonperturbative definition of ${\cal H}$ along the general
lines that we used above -- in terms of functionals of fields on a spatial
slice obeying suitable conditions -- would require a great deal of microscopic
knowledge. (What kinds of fields are considered? What are the proper
degrees of freedom to use at short distances?  What kinds of topologies
are allowed?  What are the boundary conditions near places where
the fields develop singularities?) Such a definition may not exist at all, even in principle.  
I believe that instead, by using a de Sitter analog of the familiar
holographic construction for Anti de Sitter space, we can give
a nonperturbative definition of the Hilbert space that does not depend
on any detailed microscopic knowledge (and does not make it clear
if ${\cal H}$ is finite-dimensional).  We will avoid needing microscopic
knowledge in the definitions because we will only need to know what the
fields can look like in the far past and the far future where things
are tame; we will not
need to know what sort of microscopic fluctuations are possible.

Before entering into detailed discussion, I want to point out that
this sort of approach seems to give increasingly less 
information as the cosmological constant is increased.  For negative
cosmological
constant, the sort of reasoning that we will use
gives the boundary conformal field theory, which, as discussed
above, might actually be regarded as supplying a dynamical principle.
For zero cosmological constant, we get the $S$-matrix, which at least
is a reasonable output for a theory to produce.
For positive cosmological constant, all we will really get is a definition
of a Hilbert space (unless some further meaning can be assigned
to the ``meta-observables'' that we discuss later).

Let us first recall the familar procedure for negative cosmological
constant.  We start with the Anti de Sitter metric, which near the
boundary (which we will take to be at $r=\infty$) looks like
$ds^2=dr^2+{1\over 4}e^{2r}d\vec x^2$.  We introduce an arbitrary conformal
metric $g_{ij}dx^i\,dx^j$ on the boundary and generalize the AdS metric
to one that looks like $dr^2+{1\over 4}e^{2r} g_{ij}dx^i\,dx^j$ for $r\to \infty$.
By considering the dependence on $g_{ij}$, we get the correlation
functions of the stress tensor in the boundary conformal field theory.

De Sitter space is somewhat similar but with different signature.
For $t\to \pm \infty$, the de Sitter metric of eqn.  \dsmetric\ looks like
$ds^2=-dt^2+{1\over 4}e^{\pm 2t}d\Omega^2$, with $d\Omega^2$ the
round metric on $\S^{n-1}$.
Now to prepare an initial or final state $|i\rangle$ or $\langle f|$,
we pick a conformal metric $g^{(i)}$ or $g^{(f)}$ on the sphere and
ask that the spacetime metric should be asymptotic in the far past
 to $-dt^2+{1\over 4}e^{-2t}g_{ab}^{(i)}dx^adx^b$
or in the far future to
 $-dt^2+{1\over 4}e^{2t}g_{ab}^{(f)}dx^adx^b$.

The path integral for metrics with this asymptopia in the past and
future gives an observable that we may call $\langle f|i\rangle$.
These (and their generalizations to include asymptotic fields other
than the metric)
are the only observables that I can see in asymptotically de Sitter
space time.  Actually, it might be better to call this kind of object
a ``calculable'' rather than an ``observable,'' since formulating
it requires a global view of ${\cal I}_-$ and ${\cal I}_+$ (the
infinite past and future, where $g^{(i)}$ and $g^{(f)}$ are defined), and this
is not available to any observer living in this spacetime.  So it may
be calculable, but it is not observable in the usual sense. Since
I consider the term ``calculable'' (used as a noun) to be rather clumsy,
I will instead refer to these objects as meta-observables.

For the moment, let us just try to converge on a definition of 
a Hilbert space.  The rough idea is that for any function
$\Psi(g^{(i)})$ (of a suitable class), we would regard
\eqn\plo{|\Psi\rangle=\int Dg^{(i)}\Psi(g^{(i)})|i\rangle} as a quantum state.  
However, we want to take the quotient by null vectors.
It may be the case that for some $g^{(i)}$, the matrix element
$\langle f|i\rangle$ is zero for all $g^{(f)}$.  If so, we want
to regard $|i\rangle$ as a null vector and set it to zero.

More generally, the matrix
\eqn\loko{M(f,i)=\langle f|i\rangle}
is an $\infty\times\infty$ matrix, but it may have a finite rank.
If its rank is finite, we want this rank to be the dimension of the
quantum Hilbert space.  We regard any linear combination $|\Psi\rangle$
of the $|i\rangle$'s such that $\langle f|\Psi\rangle=0$ for all $\langle f|$
as a null
vector.  Taking the quotient of the space of $|\Psi\rangle$'s by
the null vectors, we get a vector space ${\cal H}_i$ of initial states.
Likewise, taking the quotient of the final spaces by the null vectors
(such that $\langle \Psi|i\rangle=0$ for all $|i\rangle$),
we get a  vector space ${\cal H}_f$ of final states.

I have called these initial and final spaces ``vector spaces'' rather
than ``Hilbert spaces'' because as of yet, we have given no definition
of an inner product on either ${\cal H}_i$ or ${\cal H}_f$.
All we have so far is a pairing ${\cal H}_f\otimes {\cal H}_i\to \C$,
computed from a path integral for fields with a specified asymptotic
behavior in the past and future.  This pairing is bilinear (rather
than being complex linear in one argument and antilinear in the other)
since no complex conjugation is involved in computing the path integral.

By considering only the fields in the past or only the fields in the future,
there seems to be no way, without drawing on a great deal of hypothetical
microscopic knowledge, to define an inner product on ${\cal H}_i$
or on ${\cal H}_f$.  However, we can extract a Hilbert space structure
from the bilinear pairings $\langle f|i\rangle$ if we take into
account that CPT symmetry (which as far as we know is valid in the presence
of quantum gravity) gives an antilinear map from the past to the future,
that is, from ${\cal H}_i$ to ${\cal H}_f$.  If we denote the CPT
transformation as $\Theta$, then we can define a Hermitian pairing  $(~,~)$
on ${\cal H}_i$ as follows: for $|i\rangle,$ $|j\rangle\in {\cal H}_i$,
we set
\eqn\huco{(j,i)=\langle \Theta j|i\rangle.}

Since we have divided by null vectors, this hermitian product
is nondegenerate.
Now we can formulate a {\it unitarity conjecture}: it actually is positive definite.  This should be added to the {\it entropy
conjecture}: the Hilbert space defined in this fashion
is finite-dimensional, and for small cosmological constant, its dimension
is approximately given by the semiclassical entropy formula.

Note that, if our reasoning is correct, we have defined a Hilbert space
but not an $S$-matrix.  Just in order to define an inner product without
detailed microscopic knowledge, we had to use the path integral over
all of spacetime to get a pairing $\langle ~|~\rangle$ between initial
and final states.  We have no other such pairing at our disposal,
so we get a Hilbert space structure but not anything that one could
characterize as an $S$-like matrix.   

One might have been tempted to argue that the de Sitter Hilbert space cannot
be finite-dimensional, because one could in the far past divide
the space into an arbitrarily large number of causally disconnected
regions and place zero or one elementary particle in each region in
an arbitrary fashion.\foot{The notion of a particle is murky in de Sitter
space in general, but is better-defined in the far past and the far future.
Anyway, all we really need here is a modest, localized disturbance of some 
kind; it is not important to precisely interpret it in terms of particles.}
It has been argued (\banks\ and private communications by T. Banks,
R. Bousso, and G. Horowitz) that this attempt to prove the infinite-dimensionality
of the de Sitter Hilbert space fails because (given black hole
formation and the like) such a generic initial
state does not really lead to a de Sitter-like evolution.
The idea that the matrix $M(f,i)$ might have finite rank is suggested
by this argument.   Notice that from this point of view, we do not get
any insight about what would be meant by the evolution starting
from an arbitrary initial state $|i\rangle$.  All we can determine
is whether it has a nonzero pairing with final de Sitter states. 

\bigskip\noindent{\it Big Bang and Oscillatory Universes}

Defining a Hilbert space without detailed microscopic knowledge
may not seem like much, but things could have been worse.
Let us consider a Big Bang universe that is asymptotic in the future
to a de Sitter spacetime. (For a recent discussion of related
issues see \ref\bff{T. Banks and W. Fischler, ``$M$-Theory Observables For
Cosmological Space-Times,'' hep-th/0102077.}.)
 We suppose that the initial conditions
are somehow fixed quantum mechanically.  The real Universe
may be of this type.

In this case, all that we can specify is the conformal structure
$g^{(f)}$ in the far future.  The path integral with such final
conditions gives a function of $g^{(f)}$ which we might think of
as $\langle f|\chi\rangle$, where $|\chi\rangle$ is a distinguished
state of the world, determined by the quantum initial conditions.
Note that $|\chi\rangle$ is in some sense a state vector of the world,
but it is not a vector in the physical de Sitter Hilbert space ${\cal H}$
that we defined earlier.  That is because it was not produced by
specifying some incoming conformal structure $g^{(i)}$ in a world
with de Sitter-like initial conditions, but rather was generated
in some more general quantum mechanical fashion.
Since the initial conditions do not correspond to a state of the form
we considered before, we are not entitled to divide by null vectors
in the above fashion. 
The best one can say along these lines is that possibly the function
of $g^{(f)}$ given by the path integral with final conditions $g^{(f)}$
defines  a vector in an infinite-dimensional
space associated with de Sitter space.  (It is hard to give a
precise definition of this infinite-dimensional space,
 since we do not know what class of functionals $\Psi$
should be used in eqn. \plo.)  Of course the function $\langle f|\chi\rangle$
is a meta-observable, beyond the control of any observer in such a universe.

Worse from this point of view than a Big Bang (or Big Crunch) universe
would be an oscillatory universe, by which I mean simply a universe
in which we are given no statement at all about any asymptotic region
in which a simplification occurs, for example because the universe
is oscillating so that this never happens.  In this case, one could
say nothing at all without detailed microscopic knowledge. 

\bigskip\noindent{\it More On The Meta-Observables}

Now we will explore the meta-observables in somewhat more detail,
despite the fact that their physical interpretation is unclear.
The basic idea is to convert the meta-observable $\langle f|i\rangle$
into a series of correlation (or meta-correlation) functions
by expanding the conformal metrics $g^{(i)}$ and $g^{(f)}$ near the
standard round one.

To keep things simple, we will only consider the analogous expansion
for a scalar
field.  Moreover, we will set $m=0$ in \ononb, and we will set
$\alpha$ to the conformally invariant value $(n-2)/4(n-1)$.  The purpose
of this is just to keep the formulas simple; I do not believe that
this specialization affects any qualitative conclusion.

In the conformally invariant case, the propagator can be described
in a particularly simple form.  In Euclidean signature,
the  propagator between two points
$X,Y\in \S^n$ is a multiple of $1/(1-X\cdot Y)^{(n-2)/2}$.
To go to Lorentz signature, we write $X=(x_0,\vec x)$, $Y=(y_0,\vec y)$,
and then we make a Wick rotation $x_0\to ix_0$, $y_0\to iy_0$.
The propagator becomes a multiple of
\eqn\onon{{1\over (1-\vec x\cdot \vec y+x_0y_0)^{(n-2)/2}}.}
To define the meta-observables, we want to take $x_0,y_0\to \pm \infty$
and imitate the usual definition of boundary correlation functions
in Anti de Sitter space.

Since $\vec x^2-x_0^2=\vec y^2-y_0^2=1$, when $x_0,y_0\to\pm\infty$,
$\vec x$ and $\vec y$ must also diverge.  We can take
$\vec x=|x_0|\vec a$, $\vec y = |y_0| \vec b$, where $\vec a$, $\vec b$
are points in $\S^{n-1}$.  In fact, $\vec a$ and $\vec b$ are (depending
on the signs of $x_0$ and $y_0$) points
in past infinity ${\cal I}_-$ or future infinity ${\cal I}_+$; we saw
earlier that these are copies of $\S^{n-1}$.

The propagator is now for $x_0,y_0\to \pm \infty$ a multiple of
\eqn\kiho{{1\over (x_0y_0)^{(n-2)/2}}{1\over (1-\vec a\cdot
\vec b \,\,{\rm sign}\,(x_0y_0))^{(n-2)/2}
}.}  The overall power of $x_0y_0$ means, if we imitate the familiar
logic of the Anti de Sitter case, that we are dealing with a conformal
field on the boundary of conformal dimension $(n-2)/2$.
After removing this prefactor, what remains is the function
that in the AdS case we would interpret as the two point function of
a conformal field on the boundary.  It is simply
\eqn\niho{{1\over (1-\vec a\cdot\vec b \,\,{\rm sign}\,(x_0y_0))^{(n-2)/2}}.}  
One thing which is unusual compared to the AdS case is that infinity
has two components, ${\cal I}_-$ and ${\cal I}_+$, and that in \niho\
we may have a correlator between two operators both inserted on the same
component, in which case ${\rm sign}(x_0y_0)=1$, or inserted on opposite
components, in which case ${\rm sign}(x_0y_0)=-1$.  

If the two fields are inserted on the same component, we get a singularity
at $\vec a=\vec b$. This is not a surprise.  More surprising is that
for two fields inserted on opposite components, there is a singularity
at $\vec a=-\vec b$.  Though perhaps unexpected  at first sight, this 
has a simple explanation.  As we saw in analyzing the null geodesics
using the form \ormform\ of the metric,
light rays emitted in the far past at a point $\vec a$ all arrive
(independent of their direction of propagation) in the 
far future at the antipodal point $-\vec a$.  This convergence of
the light rays produces the
singularity in the ``past-future'' correlator. 

This past-future singularity is presumably related to the fact that,
classically, any initial state in de Sitter space can propagate to
a state in the far future.  If the past-future propagator were smooth
and singularity free, propagation from the past to the future
might project onto a finite-dimensional space
of states (with the other states becoming null vectors), 
as one desires to make the Hilbert space ${\cal H}$ 
finite-dimensional.  Thus, one might suspect that the past-future
correlator, after averaging over quantum fluctuations, would actually
be singularity-free.  There is actually a result in classical
relativity (pointed out in this context by R. Bousso) that goes in this
direction.  In a generic perturbation of de Sitter space, the area of
the horizon of an observer at future infinity  goes to zero
at a finite point in the past because the light rays going back
into the past have time to
converge before the Big Bang \ref\gw{Sijie Gao and R. M. Wald,
``Theorems On Gravitational Time Delay And Related Issues,''
gr-qc/0007021.}.  This contrasts with de Sitter space where, as we saw
in discussing \ormform, the horizon area is independent of time and
the backward-going geodesics only meet at past infinity, where they
produce a singularity in the past-future correlator.

There is one other important difference between the correlation functions
in the AdS case and their de Sitter cousins.  This arises because 
in the de Sitter case the spatial sections are compact and one
wants to project onto the invariants of the de Sitter symmetry group
$SO(1,n)$.  

To try to see what this means concretely, we first write down
a correlation function between $s$ fields at points 
$\vec a^{(1)},\dots\vec a^{(s)}$ in the past and $s$ fields at points
$\vec b^{(1)},\dots \vec b^{(s)}$ in the future.  
In writing the correlation function, we will for illustrative purposes
write only the terms in which all propagators connect past and future
points.  (I focus on these terms because they are the ones that
seem surprising.)  We get
\eqn\onno{\sum_\Pi \prod_{j=1}^s
{1\over (1+\vec a^{(j)}\cdot \vec b^{\Pi(j)})^{(n-2)/2}}.}
Here $\Pi$ is a permutation of the set of $s$ elements.

We have evaluated this correlator in free field theory, and of course
perturbation theory (and nonperturbative physics) will 
generate a variety
of corrections to such a formula.  But should we really expect to
get an answer of this general form, that is depending
on an arbitrary set of points $\vec a^{(j)}$, $\vec b^{(j)}$?
In a fixed de Sitter spacetime, this would be reasonable, 
but in a theory that includes gravitational fluctuations, it is not.
One has no natural way to match up past and future infinity precisely,
so one must allow for an $SO(1,n)$ rotation of the $\vec b$'s 
relative to the $\vec a$'s.
What this means in practice is that it makes sense to fix the $\vec a$'s
to arbitrary points in ${\cal I}_-=\S^{n-1}$, but then the $\vec b$'s should
be fixed only up to an $SO(1,n)$ rotation.  One way to implement this
idea is to integrate over the $\vec b$'s with a weight function
$f(\vec b^{(1)},\dots, \vec b^{(s)})$ that is 
of conformal dimension $(n+2)/2$ in each variable, so as to achieve
$SO(1,n)$ invariance.  Thus, quantities that make more sense in
the presence of quantum gravity than the correlators written in \onno\ are
an integrated version
\eqn\yonno{\int d\vec b^{(1)}\dots d\vec b^{(s)}\,f(\vec b^{(1)},\dots,
\vec b^{(s)})\sum_\Pi \prod_{j=1}^s
{1\over (1+\vec a^{(j)}\cdot \vec b^{\Pi(j)})^{(n-2)/2}}.}
It is possible to choose $f$ so that the integral converges. 
Of course, in \yonno\ we have written a free-field approximation
to the correlator.

Note that a suitable $f$ only exists for $s\geq 2$.
So in particular, such a correlator, in the presence of quantum
gravity, requires looking at the behavior of the theory at distinct
points in the infinite past and future.  So measuring such an object
is beyond the scope of any one observer in de Sitter space, who experiences
precisely one point in the asymptotic future.
In that sense, the object that we have described is perhaps better
characterized as a meta-correlator, computable and interpretable
only by an observer external to the spacetime.  This 
makes its interpretation obscure.  We turn next to the question
of what, if anything, an observer in the spacetime can measure in
a precise way.

\bigskip\noindent{\it Precision Of Physics}

We are accustomed to physical theories that make, within the
rules set by quantum mechanics, predictions of arbitrary
precision that can be tested experimentally, in principle, with
any required accuracy.  For example, we customarily assume that
the $g$-factor of the electron is a well-defined real number.
It is true that any given experiment only measures $g$ with some
limited (but perhaps very good) precision.   But one customarily assumes
 that there is
no bound to the precision with which $g$ could be measured, in principle,
given the necessary time, resources, and skill.

Likewise, any given theoretical calculation in a complicated theory
such as QED is only an approximation.  But conventionally, in flat space
quantum field theory, to the extent that our theories are correct
and thus in particular well-defined\foot{There may be a limit
to the precision with which QED can be defined
because of 
the ultraviolet behavior, so QCD, which is asymptotically free, might give a better example.}
there is no limit in principle to the accuracy
in which the computations can be done, again given sufficient resources, skill,
and patience.  We do  not assume that the theories that we have now
are valid to arbitrary precision, but we usually assume that these
theories are approximations to a better theory that does make absolutely
precise predictions, in principle, for the electron $g$-factor, the
ratios of hadron masses, and so on. 

In an eternal universe, in the absence of gravity, with a constant
free energy supply generated by stars, this makes perfect sense.
In a more realistic description of nature, taking the expansion
of the universe into account,
there are many pitfalls.

De Sitter space (or a cosmology asymptotic to it in the far future)
is a particularly unfavorable case for achieving the usually
assumed degree of precision.  For example,
if it is true that the dimension of the quantum Hilbert space is finite,
this puts a limit on the conceivable complexity of any experimental
apparatus or computational machinery.  The inflation that will occur
in the future in de Sitter space puts a limit on the time in which the
experiment must be conducted (or the computation performed) before the
free energy supply runs down.  

Even the concept of an observer in de Sitter
space as a living creature making an observation has only limited
validity.  For life itself is only an approximation, valid in the limit of a complex organism or civilization.  There might be a cosmology
in which the approximation we call life is better and better 
in the future, but this requires a process of adaptation to longer
and longer time scales and lower and lower 
temperatures \ref\dyson{F. W. Dyson, Rev. Mod. Phys. {\bf 51}
(1979) 447.}, neither of
which is possible in de Sitter space (where inflation sets a maximum
time scale, and the de Sitter temperature is a minimum temperature).
The approximation we know as life thus breaks down in the far future
in an asymptotically de Sitter world, and this will put an end
to any measurement (or computation) performed by
an observer or civilization in such a spacetime, and hence
an upper bound to its precision.\foot{I believe that for
a civilization of ever-increasing complexity to exist in the far future
also requires that the universe should have $\Omega=1$, that is, flat
spatial sections.  For $\Omega<1$, the galaxies or clusters of galaxies,
if not gravitationally bound,
recede from each other with
constant asymptotic velocities, and any civilization has only a fixed number
of atoms (or stable elementary particles) at its disposal, 
namely those that are bound to the local galaxy or  cluster.
This presumably puts
an upper bound on the possible complexity (though this last assertion
has been questioned by Dyson).  With $\Omega=1$,
it may be possible to keep absorbing matter from the surroundings and
to grow in complexity.}

It is thus just as well that the only candidates we can see for quantities
that might be calculable with arbitrary precision are the meta-observables,
which extend beyond any one horizon.  Since the horizon volume (after
a limited number of inflationary $e$-foldings) does not contain a living
observer anyway, any precisely calculable quantities associated
with the interior of a horizon would be wasted.

Where does this leave string theory?  Like physics as we know it,
string theory as we know it deals
in precisely defined quantities, such as the $S$-matrix in an asymptotically
flat spacetime, or the correlation functions of the boundary conformal
field theory for the case of negative cosmological constant.  If quantities
with this degree of precision do not exist -- which seems to be
the case in
de Sitter space if one rejects the meta-observables -- then it is not
clear just what one should aim to compute. This question has nothing
specifically to do with string theory, and any answer to it
that makes sense might make sense in string theory.

\nref\fisch{W. Fischler, A. Kashani-Poor, R. McNees, and S. Paban,
``The Acceleration Of The Universe, A Challenge For String Theory,''
hep-th/0104181.}%
\nref\suss{S. Hellerman, N. Kaloper, and L. Susskind, ``String Theory
And Quintessence,'' hep-th/0104180.}%
The problem with de Sitter space can actually be divided into two parts.
One aspect is that because of the horizon experienced by an observer,
one cannot hope to witness the final state of  the whole universe.  The other
side of the problem, which seems more acute to me,  is that, as indicated
above, one also cannot in de Sitter space make sense in a precise way of 
what we usually regard as local particle physics quantities.  
In this context, let us consider
 cosmologies that accelerate more slowly than de Sitter
space, as recently considered in \refs{\fisch,\suss}.  In the
models considered in those papers, 
the scale factor $R$ of the universe varies with 
cosmic time $t$ as $R\sim t^{1+\alpha}$ for some $\alpha>0$.
There is still an observer-dependent horizon, just as for de Sitter
space, and the final state
of the universe as a whole is not  observable.  But the curvature of
the universe vanishes for $t\to \infty$; does this mean that
local particle physics observables such as the $g$-factor of the electron
become well-defined and measurable in the far future?   Some necessary
conditions are obeyed; for example, the time scale of the cosmic
expansion gets longer and the temperature goes to zero for $t\to\infty$.
So from this point of view there seems to be no obstruction to a precise
measurement.
However, an observer in such a universe would have to perform all
experiments with a finite supply of elementary particles and free
energy stored up before the acceleration progresses too far.{}
\foot{The prospects for gathering additional matter
are less favorable than
in an open but non-accelerating universe,
which was discussed in the previous footnote,
and corresponds to $\alpha=0$.} 
Under these conditions, it seems doubtful that one could perform asymptotically
precise measurements.

\bigskip

This work was supported in part by NSF Grant PHY-9513835 and the
Caltech Discovery Fund.  I have been benefited from  explanations
by T. Banks, R. Bousso, and G. Horowitz as well as stimulating discussions
with F. Dyson.
\listrefs
\end